\documentclass[twocolumn,aps,byrevtex,showpacs,tightenlines]{revtex4}

\begin{document}
\title{Quantum coherence: myth or fact?}
\author{Kae Nemoto$^{1}$ and Samuel L. Braunstein$^2$}
\affiliation{${}^1$National Institute of Informatics, Tokyo 101-8430,
Japan\\ ${}^2$Computer Science, University of York, York YO10 5DD, UK}
\date{27 June 2003}

\begin{abstract}
It has recently been argued that the inability to measure the absolute
phase of an electromagnetic field prohibits the representation of
a laser's output as a quantum optical coherent state.  This argument has 
generally been considered technically correct but conceptually disturbing. 
Indeed, it would seem to place in question the very concept of the coherent 
state. Here we show that this argument fails to take into account a 
fundamental principle that not only re-admits the coherent state as 
legitimate, but formalizes a fundamental concept about model building in 
general, and in quantum mechanics in particular.
\end{abstract}

\maketitle

There is sometimes a clash between theorists and experimentalists
which is so deep that it seems to be based on some fundamental difference
in approach. One debate revolves around the ability to create coherent
states in real devices, e.g., in lasers. While experimentalists have been
interpreting their work in terms of coherent states for decades, some
theorists now argue that this language is invalid \cite{Rudolph01}.
A resolution to this debate is urgent not only in consideration of the last 
forty years of experimental achievements but also to provide a precise 
interpretation of present and future experimental results.

The representation of a state and its associated interpretation 
are fundamental issues. For example, in quantum theory there is
an infinite number of ensembles $\{\hat P_j\}$ for decomposing a mixed 
state $\hat \rho$ via \cite{footnote}
\begin{equation}
\hat \rho = \sum_j p_j \hat P_j\;, \quad p_j\ge 0\;,
\quad \hat P_j^2 = \hat P_j \;.
\end{equation}
Only when the state is pure does this representation become unique.
The laws of quantum mechanics say that (in the absence of any additional
information other than the state's identity) no physical 
interpretation can be based on a {\it preferred\/} choice of an ensemble
for this decomposition \cite{Kok00}. This result has been coined
the {\it Partition Ensemble Fallacy\/} (PEF) \cite{Kok00}.

For decades quantum mechanical aspects of laser science have been explained
in terms of the coherent state formalism \cite{Glauber,Sudarshan}. Recently, 
however, PEF has been used to attack the very notion of the coherent state 
in the context of continuous variable teleportation \cite{Rudolph01}. In 
these experiments coherent states were chosen as the `alphabet' transmitted 
from sender to receiver \cite{Kimble,Kimble2,Ralph}. However, if Rudulph 
and Sanders' argument \cite{Rudolph01} holds then the implications are 
significantly more far reaching than for just teleportation. In fact, 
the formalism of coherent states is a basic tool in quantum optics and 
the use of laser light to produce so-called coherent states is all pervasive.

Let us go through the argument in detail. It has long been argued, 
though without rigorous proof, that the absolute phase of an electromagnetic 
field is not observable \cite{Molmer97,Gea}. This difficulty is typically
circumvented by requiring that the nominal description of laser light as 
a coherent state $\big|\,|\alpha|e^{-i\phi}\big\rangle$ should be averaged 
over the unknowable quantity $\phi$ \cite{Rudolph01}. The resulting description 
of the laser state then becomes
\begin{eqnarray} 
\label{phi}
\hat \rho_{\strut \raise0.5em\hbox{\rm PEF}}
&=& \int_0^{2\pi} \frac{d\phi}{2\pi} Pr(\phi) \big|\, |\alpha|
e^{-i\phi}\big\rangle\big\langle |\alpha| e^{-i\phi} \big| \\
&=& \int_0^{2\pi} \frac{d\phi}{2\pi} \big|\, |\alpha|
e^{-i\phi}\big\rangle\big\langle |\alpha| e^{-i\phi} \big| 
\label{coherent_s} \\
\label{fock}
&=& e^{-|\alpha|^2}
\sum_{n=0}^{\infty}\frac{|\alpha|^{2n}}{n!}|n\rangle\langle n| \;,
\end{eqnarray}
where following Ref.~\onlinecite{Rudolph01} we have taken the prior 
probability $Pr(\phi)$ to be flat for the latter two forms. Thus, the 
ensemble of states describing a laser's output could as easily be chosen 
as a collection of number states $|n\rangle$, Eq.~(\ref{fock}), instead 
of a collection of coherent states, Eq.~(\ref{coherent_s}).
Recalling that the PEF disallows interpretations for states based on a 
preferred choice of ensemble, {\it we should infer that experiments using 
lasers cannot be reliably interpreted as demonstrating features or 
properties of coherent states}. This is the logic behind the argument of 
Rudolph and Sanders \cite{Rudolph01}.

It seems in the field of laser science that this logic has been
accepted to be technically correct, but recognized as conceptually
disturbing. Indeed it is hard to see how this difficulty would not
infect the coherent state as a general concept. However theoretical
physicists have given different reasons 
why this logic is conceptually troublesome and hence not physically
applicable. One attack is from Wiseman who although agreeing with
the argument, claims it is unacceptably pedantic, since it implies that 
we could never write down a time $t$ or a phase $\phi$ if its intrinsic 
resolution were beyond that of direct human experience \cite{Wiseman}.
A different objection has been directed towards the applicability of the 
PEF to a real laser. Here, Gea-Banacloche \cite{Gea}
and later Wiseman and Vaccaro \cite{Wiseman2} have argued that detailed
knowledge of laser dynamics {\it should\/} give extra information
about the identity of the underlying states created by a laser. This
suggests that only a preferred ensemble is physically realizable.  However, 
their analyses require the untested assumption of a perfectly Markovian 
dynamics for a laser. The Markovian assumption seems unlikely to be a 
fundamental truth. Yet another direction of attack has been made by van Enk 
and Fuchs \cite{Enk}. They claim that the actual state of a laser should be 
represented as a tensor product of repeated identical states. With such a 
restriction, the laser's state is enforced to be uniquely a coherent state. 
Unfortunately, this restriction invokes again an untestable assumption.

Given so many attempts to resolve the conflict by introduction of untestable
assumptions, let us revisit the argument of Rudolph and Sanders to
see whether it itself is free from them. In particular, the automatic 
assumption that the prior distribution of phases $Pr(\phi)$ should be 
taken as flat appears straightforward. Ordinarily, when one has an unknown 
quantity, one assigns a prior distribution based on whatever prior 
information is available. If one lacks {\it any\/} information then 
one tries to rely on symmetries in the problem. Thus, since any choice
of absolute phase $\phi$ leads to the same observable results, the flat
prior distribution appears to be the canonical choice. 

In fact, there is something fishy about this reasoning.  That a prior 
distribution is a meaningful summary of our knowledge (or lack thereof) 
depends on the full procedure of inference. Here the quantity at hand
is not simply unknown, but unknowable. If we believe the 
claim that a laser's phase is unmeasurable then no inference can ever 
be made from the prior. In other words, {\it absolutely any\/} choice of 
$Pr(\phi)$ will give identical predictions. To this extent, one's 
choice of a prior distribution for an unobservable quantity is 
a matter of `religion.' It lies outside the realm of science. 

The flip side of this argument can be found if we pick a delta function
for the prior. Such a choice reduces the density matrix to a pure (coherent)
state, thus rendering the entire application of the PEF inadmissible. 
Nonetheless, the fact remains that the two choices for the prior (flat 
or delta function) are not amenable to any physical test that will 
distinguish between them. Since the application of a principle cannot 
depend upon an untestable choice, this logic confirms our claim that the 
PEF cannot be invoked for {\it any\/} choice.

A more precise language for the states of the form Eq.~(\ref{phi}) is 
that they are cosets of operators on the Hilbert space. We are already
familiar with treating states as cosets of vectors on Hilbert space: Since 
the absolute phase $\varphi$ of a wavefunction is unobservable, all states 
$e^{i\varphi}|\psi\rangle$ are equivalent. Mathematically this coset 
structure corresponds to a projective Hilbert space.  Indeed, {\it the formation
of cosets of indistinguishable states is a universal feature of 
unobservability which induces an equivalence relation among states}.
Following tradition, we may then label a coset by any of its members. 
The realization of the underlying coset structure means that {\it any\/} 
preferred label is equally valid. This is tantamount to freedom of choice 
of a prior. For experimentalists the natural choice would then be a delta 
function, reducing to the familiar coherent-state language.

The unobservability of optical phase guarantees experimentalists the freedom 
to continue talking about a laser's output in terms of coherent states.
In fact, with the state represented in the form of (\ref{phi}) there is 
nothing to prevent experimentalists from using coherent states for
their state representation (provided any additional knowledge remains 
inaccessible). Physically then the usual coherent-state language is 
unfalsifiable. Mathematically, the freedom to choose the prior due to 
the unobservability of $\phi$ induces an equivalence relation among 
states~(\ref{phi}) over all choices of $Pr(\phi)$.

The principle that Unobservability Induces Equivalence (UIE) is, in
fact, more fundamental than quantum mechanics itself.  We would claim
that any attempt to build models about the world (quantum mechanical or
otherwise) must conform to this principle.  By comparison, PEF is only
meaningful within quantum theory. We demonstrated that the conventional 
interpretation of PEF as universally applicable is flawed.  In
particular, whenever PEF invokes inference, UIE must first be applied
to ensure that inference is possible. Thus, for the class of inference 
problems considered here the applicability of PEF is dictacted by UIE. 
It is the hierarchical ordering of principles which allows UIE to trump 
PEF. This heirarchy then allows us to pin-point the flaw in the argument 
of Rudolph and Sanders; their invocation of PEF is invalid precisely 
in the case to which they apply it: namely where a laser's phase would 
be unobservable.

\vskip 0.1truein
\noindent
SLB thanks N.\ Cohen, A.\ Peres and H.\ Wiseman for discussions. 
SLB currently holds a Royal Society-Wolfson Research Merit Award.
KN acknowledges support from the Japanese Research Foundation for
Opto-Science and Technology.


\end{document}